# Connecting Simple and Precise P-values to Complex and Ambiguous Realities

Sander Greenland

Department of Epidemiology and Department of Statistics, University of California, Los Angeles, California, U.S.A., lesdomes@ucla.edu

**Abstract**. Mathematics is a limited component of solutions to real-world problems, as it expresses only what is expected to be true if all our assumptions are correct, including implicit assumptions that are omnipresent and often incorrect. Statistical methods are rife with implicit assumptions whose violation can be life-threatening when results from them are used to set policy. Among them are that there is human equipoise or unbiasedness in data generation, management, analysis, and reporting. These assumptions correspond to levels of cooperation, competence, neutrality, and integrity that are absent more often than we would like to believe.

Given this harsh reality, we should ask what meaning, if any, we can assign to the P-values, "statistical significance" declarations, "confidence" intervals, and posterior probabilities that are used to decide what and how to present (or spin) discussions of analyzed data. By themselves, P-values and CI do not test any hypothesis, nor do they measure the significance of results or the confidence we should have in them. The sense otherwise is an ongoing cultural error perpetuated by large segments of the statistical and research community via misleading terminology.

So-called "inferential" statistics can only become contextually interpretable when derived explicitly from causal stories about the real data generator (such as randomization), and can only become reliable when those stories are based on valid and public documentation of the physical mechanisms that generated the data. Absent these assurances, traditional interpretations of statistical results become pernicious fictions that need to be replaced by far more circumspect descriptions of data and model relations.

**Introduction**

I am honored to receive the thoughtful contributions from the discussants of Greenland (2023). It appears that none disagree with my main thesis about distinguishing divergence and decision *P*-values. Furthermore, I sense or at least hope that we may be in broad agreement about the importance of distinguishing formal statistical theories and their "inferences" and "decisions" from the far more messy worlds of statistical practice, scientific inference, and policy formulation [Goodman, 2019; Toh, 2019]. Disagreements may remain however about the proper role of theoretical formalisms in real-world tasks, including how to how to account for their inevitable simplifications, how to map them into the unlimited ambiguities and complexities of real-world applications, and how to evaluate their utility in practice.

I suspect some disagreements arise largely from the different application problems we have encountered and the ways we have coped with them. Thus, in view of limited space, I will focus on the theory-application gap that inspired my paper: That, in much of science and medicine, the assumptions behind standard teaching, terminology, and interpretations of statistics are usually false, and hence the answers they provide to real-world questions are misleading. I will then turn to the illuminating examples given by the discussants, as those examples raise issues which should have been covered in detail in my paper.

**The yawning gap between theories and realities of modern statistics**

Mathematics is a limited component of solutions to real-world problems, as it expresses only what is expected to be true if all our assumptions are correct – including, especially, implicit assumptions that are omnipresent and often incorrect. The confusion of mathematical (and more generally, logical) problem solving with successful real-world application begins in basic math teaching, when students are asked to solve for the time *t* of an automobile journey given distance *d* and average speed *s* as *d*/*s* where *d* and *s* are assumed to be known exactly. In reality *d* and *s* in an actual journey will be known precisely only in retrospect. The actual time will instead depend on many prospective contingencies, such as possible alterations of *d* and *s* by detours and slowdowns. Thus, in forecasting for a decision about a route to take, we rarely have enough information to regard *d* and *s* as more than assumptions subject to what may be considerable uncertainty. One attempt to cope with that uncertainty would assign *d* and *s* a joint prior distribution based on past experience, which makes the often-costly inductive assumption that the past translates to future performance in a known fashion – useful to be sure, but dangerous if we take our model as showing all unknowns or treat our prior distribution as capturing all relevant uncertainty.

Statistical methods are rife with implicit assumptions whose violation can be life-threatening when results from them are used to set policy. An often questionable but unstated set of assumptions permeates all formal statistical methodology, including Neyman-Pearson (NP), Fisherian, likelihoodist and Bayesian of all varieties: There is human equipoise or unbiasedness in data generation, management, analysis, and reporting. These assumptions correspond to levels of cooperation, competence, neutrality, and integrity that are absent more often than we would like to believe. A classic example is that of 21 randomized trials that were reported and used to guide practice for years, but unfortunately did not exist [Rubinstein, 2009]. Such examples have been dismissed as extreme cases or labeled "hopefully rare". But thanks to the replication crisis and growing awareness of researcher degrees of freedom unaccounted for in conventional statistical interpretations [Breznau et al., 2022; Gelman & Loken, 2014; Silberzahn et al., 2018], far less extreme cases have been revealed as common, even pervasive, in medicine [Angell, 2009; Freedman, 2010; Horton, 2015].

That blatant examples of fraud took so long to be uncovered illustrates how we have no reliable meta-data on the relative frequency of such transgressions. Nor can we create such reliable meta-data given that so much basic data are either hidden from public scrutiny, impractical to access or impossible to analyze independently (the Nurses and Physicians Health Studies of Harvard's Channing Laboratory being notorious examples). Worse, as investigative sleuths have found out, journals are loathe to admit and retract studies in which fatal errors were discovered, and the literature at large may long continue to cite such studies as if no such discovery occurred [Doshi, 2015; Allison et al., 2016].

In light of this harsh reality, we should ask what meaning (if any) can we assign to the *P*-values, "statistical significance" declarations, "confidence" intervals, and posterior probabilities that are used to decide what and how to present (or spin) discussions of analyzed data. The standard interpretations are couched in terms of error rates and rational bets, each of which condition on the correctness of all assumptions used to derive those rates or bets: I call the set of those assumptions the target model *M*. That set includes assumptions that are usually unstated and often violated, such as no unaccounted-for selection biases, data errors, or data dredging (whether searching for p≤0.05 or for p>0.05). My colleagues and I have long argued that inferences and decisions that pivot on formal statistics are misleading when the application context leaves us uncertain about those assumptions [Greenland, 2005, 2017; Greenland & Lash 2008; Rafi & Greenland, 2020; Amrhein & Greenland, 2022; Greenland et al., 2023]. In those contexts, careful uncertainty analysis will reveal that the stated "significance" and "confidence" levels reflect only naivety about the assumptions needed for those levels to hold [Greenland, 2005; Greenland & Lash, 2008], so that placing 95% confidence on a nominally 95% "confidence" interval is overconfidence [Amrhein et al., 2019a, 2019b; Greenland, 2019a, 2019b].

**Unconditional measurement: Harm reduction without revolution or complete solution**
In the face of doubtful assumptions, we have called for teaching, terminology, and interpretation of statistics to switch to *logically unconditional* forms. Here, "unconditional" means that the re-interpretation of the statistic remains a correct observation *even if the assumptions used to derive the statistic are incorrect* [Rafi & Greenland, 2020; Greenland et al., 2023]. An everyday example of due caution is to say "my clock says it's 4pm", which is a description of an observation and allows that the clock may be incorrect, but does not purport to supply the error rates of the clock (e.g., from failure to set the clock forward upon change to daylight savings). We usually compress the statement to "it is 4pm", which is a claim about the actual current time, not a clock observation. This compression implicitly assumes that the clock is correct; its use thus conditions on high confidence that the clock was set properly and is running within acceptable error rates – mechanical assumptions which, when violated, could lead to missing a plane or train.

Standard interpretations are akin to saying "it's 4pm" without caution: Claiming an error rate of α for a "*p*≤α" decision or "inference" rule for rejecting *M* assumes that only narrow violations of *M* are possible – namely, violations of the set of assumptions H that *M* includes beyond the set of background assumptions *A* (the embedding model). The description may however become ever more irrelevant if not misleading about the target if the embedding model *A* is itself far from correct (describes the behavior of the data generator poorly), as signaled for example by a large divergence of the data (considered as a point mass in distribution space) from the family of distributions that obey *A*. Even if there is no questionable assumption about the research conduct, this concern can arise when, as often the case, *A* contains typical oversimplified parametric constraints (such as linearity assumptions). Such incorrect constraints drive the data projection onto the *A* family away from the data, to the point where it may be called oversimplification or oversmoothing. Such oversimplification is aggravated when *A* (rather than *M*) is chosen by standard data-driven model simplifications, as when an analysis targeting a treatment coefficient in a regression model begins with data-based selection of covariates.

Some limited protection against oversimplification is obtained by keeping *A* very weak (small), as in preliminary data smoothing [Greenland, 2006]. But in practice analysis more often begins and ends with an oversimplified model, and we are confronted with uncertainty not only about H but also about *A*. In that case we should know what our statistics mean unconditionally, regardless of the correctness of *A* or *M*. An example is to say that a Fisherian *P*-values describes a divergence between two data projections: One projection is under a weaker (smaller) assumption set *A* that defines the model in which the analysis is embedded, such as a flexible regression model; the other projection is under a stronger (larger) assumption set *M* that the analysis targets, such as a submodel defined by adding the assumption to *A* that a targeted coefficient has a known value (usually zero).

This unconditional description does not assume that either model is correct, and does not supply the error or coverage rates of (say) a decision rule under all conceivable violations of *M*, such as those that also violate *A*. The context determines the manner and extent to which a divergence statistic remains relevant or useful; it might be judged relevant even if *M* is known to be incorrect, as when *M* is recognized as no more than an approximation to reality [Box, 1980], or when *M* refers only to a counterfactual world, as in formal causal analyses where *M* could have been forced to hold but wasn't [Hernán & Robins, 2020; Pearl, 2009; VanderWeele, 2015].

In sum, *P*-values may be helpful for summarizing how well a model in the family defined by *M* can describe the data or some simplification of the data based on *A*. That does not however make them anywhere near sufficient or even necessary for that description, let alone safe for mandated use. All this seems consistent with views expressed, for example, in Lavine [2019], allowing that a *P*-value is but one very specific measure for judging the divergence of a model *M* from the observed data or the data smoothed under the assumptions in *A*.

**Ordinal vs. Metric Information**

Prof. Lavine remarks in a footnote that I seem "to use descriptive, discrepancy, and divergence interchangeably to describe the same type of *P*-value". I should have clarified that I use the terms "descriptive" and "discrepancy" to refer to *P*-values that can be interpreted as single-sample summaries of relations of models to data or to one another. That includes for example *P*-values from rank-sum statistics, which ignore spacing (metric) information on individuals when measuring discrepancies between distributions. I use "divergence" more narrowly, to refer to *P*-values derivable from geometric-divergence statistics which employ that information [Amari, 2016]; those include everyday statistics in parametric and semi-parametric modeling such as sum-of-squared-deviation (SSD) and likelihood-ratio statistics. A divergence *P*-value is then an ordinal description of a geometric measurement.

That said, I am puzzled by Prof. Lavine's objection to taking ordinal measures as one type of evidence summary. We use ordinal evidence measures in everyday life. This happens, for example, when an employer hires an applicant in part based the applicant was in the top 10% of their graduating class by some measure. Here, an ordinal ranking is being taken as evidence for future performance superior to those of lower ranking. We know such measures are far from sufficient for real decisions, and that rankings discard spacing information; the same is true of *P*-values. But they do retain *ordinal* information.

In particular, a *P*-value contains and an *S*-value measures the refutational information given by a ranking of the observed statistic in a distribution. That is certainly not all the information in the data about the targeted hypothesis or model. Nonetheless, as any nonparametric statistics book shows, rankings do contain information that may be taken as relevant evidence regarding differences among distributions. Of course, the spacing information discarded by rankings is also important, but it is more intelligibly captured by estimates. Note however that the use of spacing information may incur a considerable increase in risk of bias due to violations of background assumptions, as when $M$ entails a dimension reduction or model simplification relative to $A$ such as setting a coefficient to zero.

**Why divergence statistics cannot provide support without restrictive assumptions**

Prof. Gasbarra is concerned that divergence can only serve as evidence against $M$, never in support of $M$ unless $A$ severely restricts alternatives to $M$. I am puzzled by that concern in part because Prof. Gasbarra mentions Popper, who explained at book length why, without what may be unacceptable restriction, evidence can only refute theories (e.g., Popper [1959, 1962]). The need for a restrictive $A$ in measuring support is implicit in pure likelihood (e.g., Edwards [1992]; Royall, [1997]), where support is only relative to alternatives, and in pure Bayesian methods, where a formal and sharply circumscribed possibility space is needed to define a prior distribution. There are always alternatives to unsaturated models that are far better supported by the data in pure likelihood terms, including but not limited to the possibility that the data were constructed to produce exactly what was observed (e.g., Rubenstein [2009]). Without restrictions in $A$ to exclude such alternatives, discrepancies from what $M$ would lead us to expect can only serve as refutation measures, as Karl Pearson and R.A. Fisher treated $P$-values. Adding the restrictions needed to build sensible support measures to $A$ will drive $A$ and thus the analysis further from the data, increasing the risk of oversimplification.

**Benefits and hazards of narrowing statistical alternatives to enhance senstivity**

Any one-dimensional summary statistic for a multidimensional concept (such as discrepancy) must discard much information about lack of model fit – that is, refutational information. This dimension reduction will limit the types of discrepancies the statistic will be sensitive to. For example, in assessing departures from a no-association model, the usual linear-trend $P$-value will be sensitive to linear components of associations, but insensitive to nonlinear components [Maclure & Greenland, 1992]. As such examples show, and Professor Rice notes, to say that $P$-values need involve no alternative is a mistake, one which I have fallen prey to at times: Our choice of statistic should indeed depend on what alternatives we want sensitivity for. Narrowly targeting precise alternatives will however leave us with statistics insensitive to other alternatives, and optimization for the targeted alternatives can worsen that insensitivity to other important violations of our models.

*P*-values are but preliminary checks or diagnostics for *something* out of line with what the model assumed. Here we encounter the No-Free-Lunch-Principle (NFLP; van Zwet et al. [2021]): The more sensitive we make a diagnostic to one class of alternatives, the less sensitive we make it to another. Making *A* more specific to encompass only precise, limited statistical alternatives to *M* risks the "small world" trap: Conclusions from comparisons within a highly constrained world (a restrictive *A*) can be completely reversed under alternatives outside that world. Again, this can be seen in simple examples in which *A* allows only linearity but reality may be nonlinear [Maclure & Greenland, 1992], as well as in analyses of ordinarily unmodeled uncertainty sources, such as errors in patient recruitment and data management which are likely to go undetected when they lead to no clearly suspicious data pattern.

Prof. Gasbarra presents an example of this problem involving a common *P*-value: The upper tail area from the usual $\chi^2$ fit statistic (which is a squared standardized distance of a data vector from its expectation). He notes how this *P*-value is completely insensitive to overfitting and manipulating data to favor *M*, thus necessitating attention to the lower tail. I mentioned this issue in the last paragraph of sec. A.8 of my paper. I am grateful that Prof. Gasbarra has afforded me the opportunity to elaborate on what was clearly an insufficient discussion of how the lower tail becomes relevant.

First, one can view statistical overfitting as an analysis problem of using a reference distribution that does not match the distribution of the statistic actually used. The ordinary upper-tail $\chi^2$ *P*-value assumes overfitting is absent, thus making that absence an assumption in *A* and rendering the *P*-value completely insensitive to underdispersion in general and overfitting in particular. If that overfitting arises from selecting the statistic by a programmable (algorithmic) selection process, the problem can be addressed by simulation of that process (e.g., bootstrapping) to approximate the actual distribution. The overfitted statistic is often a minimum of $\chi^2$ statistics over whatever set of models was examined; it is thus shifted toward zero relative to a single $\chi^2$ distribution, and its distribution can be simulated if we know set of candidate models and the selection criterion.

In contrast, if we are not given enough information to simulate the model-selection process, or the problem is data manipulation, the most we can do may be to look for signs such as underdispersion. In these cases the lower as well as upper $\chi^2$ tail may be seen as arising from data sets far from the bulk of the data distribution ("noise cloud") implied by *M*. More generally, divergences quite different from squared distance can also be essential when we confront nonregular distributions, as will be discussed below regarding Prof. Lavine's examples.

**Should fit statistics default to 2-tailed *P*-values?**

When we spot a discrepancy well beyond what our model would lead us to expect, we can and should turn to causal explanations for the discrepancy. A classic example Fisher's use of the lower tail of a $\chi^2$ statistic as a diagnostic tool [Fisher, 1936, p. 129-132]. Suspicions had been raised that Mendel's data had been manipulated to fit expected values under Mendel's theory. Those causal suspicions dictated the direction to examine deviations beyond the noise expected under the theory, as that noise all but guaranteed substantial departure from those expected values. Thus, upon finding a very small departure, Fisher wrote that "there can be no doubt that the data from the later years of the experiment have been biased strongly in the direction of agreement with expectation."

The motivating suspicions might never had been raised had there been trusted documentation of data collection and management that supported putting the assumption of "no experimental bias" in *A*. In that case the small lower-tail *P*-value could have been dismissed as an improbable but real chance deviation, or perhaps indicative of some unknown natural source for the underdispersion. Thus, the traditional use of only the upper tail implicitly includes "no underdispersion" (whether from bias or natural phenomena) in *A*.

I would argue that the upper-tail-only default is incorrect when underdispersion is not assured by design and context. In that case the underdispersion assumption should be seen as part of H rather than *A*. With 3 or more degrees of freedom, this means that the $\chi^2$ *P*-values for fit should include both tails to provide sensitivity to underdispersion. Geometrically, this treatment follows from viewing the data distributions under *M* in the *n*-space of possible data vectors (the full sample space) and their image in the range of their $\chi^2$ statistics; there it can be seen that both tails are composed of statistics beyond the bulk of the distribution under *M*. (With only 1 or 2 degrees of freedom, the mode of the $\chi^2$ statistic is zero, hence the data would have to be disaggregated to provide more degrees of freedom for sensitivity to underdispersion.)

**Sensitivity, power, cutoffs, and statistical blindness to causal alternatives**

Fisher's example illustrates why much more than a *P*-value or CI is needed for anything beyond mere description of a data-model relation: We need a contextual causal narrative to see clearly what should and shouldn't be assumed as background when computing and interpreting the statistics [Greenland, 2022]. I align with views that the imposition of sharp cutoffs and declarations based on them, as central to Neyman-Pearson theory, can be destructive and misleading. Conformity to that theory has led to redefinitions of "*P*-value" that are divorced from concepts of divergence measurement and that introduce incoherencies. Worse, the assumptions needed for Neyma-Pearson interpretations increase vulnerability to the "small world" trap, at least outside of ideal experiments that create the assumed small world in the experimental reality. Such ideal experiments are far more rigorous and error-free in design, execution, and reporting than are most actual clinical trials [Angell, 2009; Horton, 2015].

To repeat: In real applications there are always alternatives to which a *P*-value will be sensitive, others to which it will be blind, and examples of each type that are vague or unrecognized, such as fraud and innocent procedural error. That reality is what I had in mind in saying that *P*-values do not assume or demand precise alternatives. But in NP decision theory, sensitivity is replaced by the much more precise design concept of power, and that requires far more extensive, precise specification. Such precision is usually misleading because of gross uncertainties about both the specified parameters and its implicit background assumption that all possible model departures have been accounted for. Worse, as jaded research-proposal reviewers know all too well, claimed power is typically biased upward by specifications chosen to inflate power claims and thus motivate funding.

Furthermore, power calculations are tied to fixed *P*-cutoffs or α-levels, whose problems for assessment of specific hypotheses are legion but widely ignored in practice. On this issue, Royall (1986) concluded that

> "the degree of creativity and sensitivity to semantic nuances that is required [to properly interpret *P*-values as evidence] seems incompatible with the popular view that the significance test is a commonsensical, practical tool appropriate for wide-spread use in scientific reporting."

That lament remains true [Goodman, 2016], but given the unstoppable ubiquity of *P*-values and functions of them, we should accept our responsibility to supply the nuances. One way of doing so is to show how *P*-values can and should be freed from the confusion caused by constantly wedding them to controversial additions like sharp cutoffs, "statistical decisions," "significance" declarations, and "statistical hypothesis tests", and should be returned to their original form as continuous descriptors of contextually relevant discrepancy measures [Amrhein et al., 2019a, 2019b; Amrhein & Greenland, 2022; Greenland et al., 2016, 2022, 2023; Hurlbert & Lombardi, 2009; McShane et al., 2019; Rafi & Greenland, 2020; Rothman et al., 2008, Ch. 10; Wasserstein & Lazar, 2016; Wasserstein et al., 2019].

Once we are free of the overinterpretation of p=0.05 as "significant" in the sense of declaring a model or hypothesis has been refuted, the evidential or information content of common *P*-values may be seen as slight both via sampling experiments [Gelman & Stern, 2006] and by direct information rescaling to surprisals (*S*-values): $p = 0.05$ is nothing more than a rough border between negligible and not-quite-negligible, hardly different from getting all heads when tossing a coin 4 times to judge fairness [Rafi and Greenland, 2020; Cole et al, 2021; Amrhein & Greenland, 2022; Greenland et al., 2022; Heumann et al., 2022]. At most, then, a modestly small *P*-value like 0.05 may call for a closer look and possible replication if worth the effort on contextual grounds. I think Fisher intended 0.05 that way and was incensed by its transmutation into a natural constant like $\pi$ or *e*: In an anecdote told to me by H.O. Lancaster, when Fisher was asked if he regretted anything in his illustrious career, he snapped back "Ever mentioning 0.05!"

Particle physics learned this lesson the hard way as it saw *P*-values much smaller than 0.05 lead to no replication in time-consuming and expensive experiments. By the time of the celebrated Higgs-boson detection, it was demanding a *P*-value of at most 1 in 3.5 million (the normal tail area above 5σ) in **two** independent experiments as necessary to declare discovery, *even though prior beliefs overwhelmingly disfavored the model M targeted by those P-values* (which was the standard model without the Higgs boson). This demand has been described as crude protection against modeling mistakes and other so-called systematic effects that were either unknown or not properly taken into account [Cousins, 2017, p. 425].

The general problem addressed by this severe solution can be framed as the blindness of statistics (whether *P*-values or posterior probabilities) to differences among causal explanations for observations: If a detection event has a small *P*-value, that statistical observation cannot by itself distinguish the detection caused by the targeted phenomenon from spurious detection caused by uncontrolled observational artefacts; and if the event has a large *P*-value, that statistical observation cannot distinguish absence of the phenomenon from nondetection due to uncontrolled artefacts. Only physical prevention of artefacts can enable us to move beyond statistics to scientific inference.

**Like all statistics, *P*-values are only pieces of evidence**
Regardless of the observed magnitude of a *P*-value, Prof. Lavine's astute comments reveal that I should have emphasized how any statistic, such as a *P*-value or likelihood ratio, is only a *piece* of evidence, and often a minor one at that [Amrhein et al., 2019a, 2019b; McShane et al., 2019]. The practical significance of such statistical fragments has been blown far out of proportion, diverting attention from crucial components of *physical* evidence that have ghostly representations in theory via abstract mathematical restrictions. The evidential content of quantifiable physical restrictions can be measured in various ways, such as via increase in Fisher information or reduction in estimator variance [Seidenfeld, 1992]. In particle-physics experiments, these evidence components are documented in extensive mechanical checking and calibration of equipment to eliminate "loopholes" that could produce a small *P*-value even if the targeted model *M* were perfectly correct, or produce a large *P*-value even if *M* was very wrong. In either case, the crucial *scientific* evidence comprises physical assurances for the strong background assumptions needed to support conventional statistical interpretations.

When such physical evidence is undocumented or suspect, the *absence* of such assurances should be accommodated by switching to more modest descriptions of evidential relations, as in changing from "significance" and "confidence" to weaker compatibility language [Greenland, 2019ab; Rafi & Greenland, 2020; Cole et al., 2021; Amrhein et al. 2019a, 2019b; Amrhein & Greenland, 2022; Greenland et al., 2022, 2023]. A major obstacle to adoption of this reform is that neither statisticians nor researchers want to face how weak all statistical evidence is when the assurances are absent or suspect. Instead, most reasoning in medical research reports proceeds as if the data supply far more information than they do. This is seen in how medical research reports remain dominated by "significance" and "nonsignificance" declarations, or equally naïve descriptions based on inclusion in or exclusion from "confidence" or posterior intervals, despite the numerous uncertainty sources those declarations and intervals neglect.

At best such declarations are followed by sensitivity or bias analyses to provide some accommodation to assumption uncertainty. Unfortunately, there is a vast number of ways that the background assumptions can be varied, which leads to many researcher degrees of freedom in choices for sensitivity analyses. As in conventional analyses, simplifications to reduce the dimensionality of the choices are typically made via independence assumptions, which often turn out to be absurd in contextual terms [Greenland, 2001; Greenland and Lash, 2008]. More generally, the validity of sensitivity analyses hinge on assumptions of expertise, competence, neutrality, and integrity that are themselves often in doubt [Maclehose et al., 2021]; even when those assumptions are satisfied, there are often far too many uncertainties about more basic assumptions to analyze within reasonable time [Greenland, 2021].

In light of these problems the weaknesses of conventional terminology seem profound. For example, if the sample size is made enormous enough (as in "big data" studies), ordinary "confidence" and posterior intervals will narrow around the point estimate. Yet even modest assumption violations (systematic errors or unmodeled bias sources) will offset that narrow interval from the true parameter value and thus make it probable the interval excludes reality. Thus the confidence coefficient of the interval will represent false confidence, and any bet favoring the interval will entail almost sure loss – despite the interval being calibrated and the bet being coherent *under the assumptions in their derivations*.

Objections to switching from "significance" to "incompatibility" and from "confidence" or "credibility" to "compatibility" appear blind to these problems. Whatever convention is used for "highly" (e.g., 95%), it is far more cautious to say that, *by the measure used*, "the values in the interval appear highly compatible with the data" than to say that "we are highly confident" or "it is highly credible" that the true value is in the interval. As a consolation, the switch to "compatibility interval" leaves the interval acronym "CI" undisturbed.

Even in this limited, cautious role as compatibility measures, I think traditional statistics like *P*-values are best transformed away from the probability scale (where they are inevitably confused with posterior probabilities) to more equal-interval refutational scales such as surprisals. I thus am most pleased to see Prof. Bickel's extensions to aid that reform.

**The examples in Prof. Lavine's commentary and their implications**

This section is more technically detailed than the rest of the present paper, and requires access to Prof. Lavine's comment; it can be skipped without loss of continuity. Nonetheless, Prof. Lavine's examples are quite instructive. Although not stated explicitly, they appear to take the squared distance of a scalar data point *x* (*n*=1) from the mean of 0.5 under *M* as the divergence statistic described by the *P*-value, with no restriction on alternatives to *M* other than that the data distribution is on the unit interval; thus the model space defined by *A* comprises all distributions on that interval. Example 1 may be controversial, as one could argue that the mean-divergence *P*-value p($x_2$) ≈ 0.1 is picking up the following discrepancy missed by the unrestricted likelihood ratio: In the space of distributions on the interval, *M* is not very close to the distribution with point mass at $x_2$ = 0.945 when we take a squared distance between means as our divergence measure.

A mean divergence would be especially relevant if *A* contained a further restriction to the unimodal families commonly used in practice. But example 2 rightfully reminds us that, despite a high *P*-value, a divergence statistic can fail to capture how the observed data can be highly incompatible with the target model *M*, as seen with other divergence statistics. Specifically, if *M* is as in Lavine's Fig. 2, a mean divergence is a woefully insensitive measure for unspecified departures from *M*, since models that depart radically from *M* may have the same mean but be far away from *M* under other measures; Lavine's Fig. 1 is an example.

As discussed in sections 2.2, 4.3, and A.3-A.5 of my paper, using the Kullback-Liebler divergence between distributions translates into using negative-log likelihood ratios or deviance differences as divergence statistics, which can be can be applied beyond the scope of mean divergences. Thus, to address example 2 within its unrestricted embedding model *A* (which has zero deviance), as in sec. 2.2 in my paper, consider the negative maximum loglikelihood under *M*, or twice that, the *M*-deviance dev(*M*), as the divergence measure. Under the *M* in Fig. 2, the exact distribution of dev(*M*) yields *p* = 0.01(0.8) = 0.08 for any *x* between 0.1 and 0.9. I think *p*=0.08 is in accord with Prof. Lavine's assessment that *x* = 0.5 is not very compatible with *M*. Nonetheless, that *p* translates to only $-\log_2(0.08) \approx 3.6$ bits of information against *M*, less information than in 4 coin tosses to check fairness; but then the data comprise only one draw from *M*, and the *P*-value distribution from dev(*M*) is far from uniform under *M*: It is binary with mass 0.08 on *p*=0.08 and 0.92 on *p*=1.

That the deviance *p*=0.08 while the mean-divergence *p*=1 illustrates a key point: A large divergence *P*-value only tells us the divergence measure it describes is within its probable range under the model, so that *M* appears compatible with the data *by this measure*. Without restrictions in *A* that make a measure sufficient for describing compatibility, a large *P*-value for that measure is far from any guarantee of compatibility under other measures. In the same way, with no restriction on the set of pants in which we search, finding that a subset of pants is compatible with our leg length is no guarantee the pants are compatible with our body (fit well enough to use), as many older adults have discovered when trying on clothing from their young adulthood. Leg-length compatibility might however suffice if we could restrict our search to a set of pants compatible with our waist size, assuming we regarded other aspects of fit as ignorable.

Example 3 further illustrates the insensitivity of mean divergences to discrepancies that aren't reflected in mean shifts, and raises some subtleties about transforms of random variables. In both example 2 and 3 the mean divergences are zero and hence both their *P*-values are 1. But the models for the distributions of the random variables *X* in example 2 and *Y* in example 3 (which are what the frequentist statistics refer to) have vastly different shapes, as seen by comparing Figs. 2 and 3. This shape shift reflects how *Y* is a very nonlinear transform of *X*, as seen in Fig. 4. That the mean 0.5 happens to be a fixed point of the transform makes E(*Y*) = E(*X*), but does not make the compatibility of *x*=0.5 with Fig. 2 equivalent to the compatibility of *y*=0.5 with Fig. 3. The compatibility difference is reflected by the exact *P*-values derived from dev(*M*), which are *p*=0.08 in example 2 and *p*=1 in example 3, both of which seem reasonable compatibility descriptions for observing 0.5 upon single draws from their very different models.

The illustrations in my paper excluded Lavine's examples by taking means or mean-regression coefficients as target parameters and then including sufficient regularity conditions in *A*, as would be induced by fixed sampling from standard exponential-family GLMs and their semiparametric extensions. Such restrictions subsume the vast majority of statistical methods in health and medical research, although they are often criticized in philosophical discussions. With those restrictions, *P*-values from standardized mean divergences have the same first-order local asymptotic behavior (including optimality properties) as the corresponding likelihood-ratio statistics, which avoids apparent conflict with pure-likelihood descriptions (as noted by Royall [1997, Ch. 6]).

In sum, the lesson I see in Lavine's examples (and neglected in my paper) is that the analyst needs diagnostic statistics whose distributions are sensitive to model violations of concern (e.g., Cox and Hinkley, 1974, Ch. 3; Cox, 1977). Furthermore, in most applications, adequate model evaluations will require multiple statistics. To address multiplicity concerns those statistics could be combined into a multidimensional summary; the Pearson $\chi^2$ statistic is a prototypic example, as it sums the divergence of every cell in a contingency table. More generally, nonregular examples deserve emphasis for showing that any scalar diagnostic (including a *P*-value) measures but one aspect of compatibility or incompatibility or refutation, and will have limitations that need to be borne in mind. I should have elaborated on that fact in my paper, and am grateful to Prof. Lavine for prodding its discussion.

**Some general responses and recommendations**

It is fair to ask: Why bother with all this? Given generations of controversy, why not just scrap *P*-values? One reason is that all formal statistics are subject to the most serious objections used to attack *P*-values, such as oversimplified and misleading interpretation, ritualistic use, vulnerability to gaming, and contextual irrelevance [Amrhein et al., 2019a, 2019b; Benjamini, 2016; Gelman, 2016]. There is hope that competent, knowledgeable analysts will reach similar conclusions about what data show with whatever methodology they use [Dongen et al., 2019] – albeit the examples displaying such agreement are relatively free of the interacting cognitive biases and conflicts of interest that dominate medical controversies [Angell, 2009; Greenland, 2017].

Even if one wants to drop *P*-values from their own work, one should recognize that *P*-values appear to be here to stay despite calls for their prohibition. Thus, we must learn to live with *P*-values and their summaries such as CIs even as we campaign against their degradations such as fixed-cutoff significance tests [Goodman, 2016, 2019; Greenland and Poole, 2013a, 2013b; Rothman et al. 2008]. As an analogy too close for comfort, consider that, after decades of concerted resistance by tobacco interests and some heavily invested physicians, smoking was finally recognized as a leading source of life-threatening diseases. Rather than repeat the policy mistake of outright prohibition (as in the disastrous U.S. experiment with alcohol in the 1920s), tobacco smoking was reduced through steady education efforts, including deglamorization of cigarettes and development of less harmful replacements that provide directly measured nicotine without inhalation of toxic combustion products.

Similarly, despite decades of concerted resistance by those heavily invested in significance testing and NP decision theory, many scientists and statisticians have recognized that identifying *P*-values with "statistical significance levels" and interval estimates with "confidence intervals" are leading sources of distorted research reporting [Amrhein et al., 2019a, 2019b; Greenland, 2019a, 2019b; Greenland et al., 2022, 2023; McShane et al., 2019; Rafi & Greenland, 2020; Rothman et al., 2008; Wasserstein et al. 2019], with the most prevalent distortion being overstated certainty in conclusions – "uncertainty laundering", as Gelman [2016] put it. We need to deglamorize so-called "test statistics" by demoting them from truth indicators to limited evidence summaries. We also need to teach how decision rules and support measures can be seriously misleading when computed from doubtful assumptions; such methods thus should be left to analyses of tightly controlled experiments that physically enforce those assumptions.

In many research reports, this conceptual harm reduction may change neither computation nor numerical content. But it will emphasize the insufficiency of statistical summaries for real-world inferences and decisions, shifting focus to documented design features and to the field efforts to enforce those features. Such efforts are the ultimate source of real-world (as opposed to hypothetical) information, via the assumptions they mechanically force to become physical reality [Greenland, 2022]. To enable this conceptual reform, we must provide modest framing of statistical summaries as descriptions of observations in relation to models and theories, without the intoxicating intellectual smoke embodied in terms like "statistical inference" and "statistical decisions". That reform will require teaching the distinctions between descriptive and decision *P*-values and CIs, and explaining why most research reality can justify these statistics only as imperfect evidence measures unworthy of error-rate claims derived from calibrations within unrealistically small worlds.

Finally, it should be expected and taught that no multidimensional real-world concept (whether significance, confidence, compatibility, coherence, refutation, or pants size) can be captured adequately by a single scalar measurement. That is one reason to not promote *P*-values over other diagnostics. Other diagnostics have their own limitations, however. For example, pure Bayesian diagnostics need embedding assumptions so narrow and specific to enable one to define a contextually intelligible prior distribution over all distributions allowed by *A*. This limitation that has at times been recognized and answered in Bayesian model checking by frequency-calibrated *P*-values [Box, 1980; Bayarri & Berger, 2000; Robins et al. (2000)].

**Confronting cognitive biases**
I have come to think that many statistics books and instructors as well as users fail to understand vital aspects of interpreting *P*-values, α-levels, and CIs. As I have written at length elsewhere [Greenland, 2017], I believe these misunderstandings stem from a set of interrelated cognitive biases that reflect innate human compulsions which even the most advanced mathematical training seems to do nothing to staunch, and may even aggravate: Dichotomania, the tendency to reduce quantitative scales to dichotomies; nullism, the tendency to believe or at least act as if an unrefuted null hypothesis is true; and statistical reification, the tendency to forget that mathematical arguments say nothing about reality except to the extent the assumptions they make (which are often implicit) can be mapped into reality in a way that makes them all correct simultaneously.

I find that these biases are usually unstated or denied, yet permeate the research literature and statistics tutorials. They synergize to produce dramatic understatements of uncertainties about the hypotheses, models, and parameters the methods purport to address, resulting in vastly overconfident real-world interpretations of statistical methods and their outputs. I have received extensive agreement with this harsh assessment in feedback from applied scientists of high intellectual integrity, skill, knowledge, and experience in their fields (some of whom have graduate statistical degrees, even doctorates).

It would not be an exaggeration to say that many of us see academic statistics as "the sick man of science," an alarming diagnosis in light of the central role it plays in science in general. The situation is maintained through specious and often byzantine rationales for destructive traditions, which is as expected from another human condition: Blindness to conceptual mistakes when correcting those would threaten the entire foundation on which one's teaching and research (which is to say, entire careers) have been built. The central example is how, by themselves, *P*-values and CI do not test any hypothesis, nor do they measure the significance of results or the confidence we should have in them. The sense otherwise is an ongoing cultural error perpetuated by large segments of the statistical and research community via evocative terminology such as "statistical significance" and "confidence interval" for what are only dichotomous *P*-value summaries.

In summary, a divergence *P*-value is an ordinal summary of a measurement of the relation between data and a model for the behavior of the actual data-generating process. It is a straight (even if not straightforward) description about where the measurement fell in a hypothesized distribution. Such distributions and hence *P*-values and CIs can only become contextually interpretable when derived explicitly from causal stories about the real data generator (such as randomization), and can only become reliable when those stories are based on valid and public documentation of the physical mechanisms that generated the data. Absent these assurances, traditional interpretations of "inferential statistics" become pernicious fictions that need to be replaced by far more circumspect descriptions of data and model relations.

**Acknowledgement**: I am grateful to Valentin Amrhein for helpful comments on the draft of this rejoinder.